\documentclass{article}%
\usepackage{amssymb}
\usepackage{amsfonts}
\usepackage{amsmath}
\usepackage{graphicx}%
\usepackage{fullpage}
\newtheorem{theorem}{Theorem}

\newtheorem{definition}[theorem]{Definition}

\newtheorem{proposition}[theorem]{Proposition}

\newenvironment{proof}[1][Proof]{\noindent\textbf{#1.} }{\ \rule{0.5em}{0.5em}}

\newcommand{\FirstName}{\mbox{FirstName}}
\newcommand{\LastName}{\mbox{LastName}}
\newcommand{\Neighbour}{Neighbour}

\newcommand{\e}{e}
\renewcommand{\c}{c}
\renewcommand{\t}{t}
\newcommand{\name}{name}
\newcommand{\C}{C}
\newcommand{\D}{\mathcal{D}}
\newcommand{\F}{F}
\newcommand{\G}{G}
\renewcommand{\L}{L}
\renewcommand{\P}{P}
\newcommand{\Q}{Q}

\renewcommand{\S}{S}
\newcommand{\T}{T}
\newcommand{\A}{A}
\newcommand{\X}{X}
\newcommand{\Y}{Y}
\newcommand{\V}{V}
\newcommand{\I}{I}
\newcommand{\SN}{SN}
\newcommand{\male}{\mathit{male}}
\newcommand{\female}{\mathit{female}}
\newcommand{\dom}{dom_{\mathcal{D}}}
\newcommand{\con}{con}
\newcommand{\supp}{support}
\newcommand{\apriori}{\textsc{Apriori }}
\newcommand{\warmr}{\textsc{Warmr }}
\newcommand{\tuples}{tuples_{\mathcal{D}}}

\begin{document}

\title{Association Rules in the Relational Calculus}
\author{Oliver Schulte, Flavia Moser, Martin Ester and Zhiyong Lu\\School of Computing Science\\Simon Fraser University\\Burnaby, B.C., Canada\\\{oschulte,fmoser,ester,zhiyongl\}@cs.sfu.ca}
\maketitle

\begin{abstract}
One of the most utilized data mining tasks is the search for association
rules. Association rules represent 
significant relationships
between items in transactions. We extend the concept of association rule to
represent a much broader class of associations, which we refer to as
\emph{entity-relationship rules.} Semantically, entity-relationship rules
express associations between properties of related objects. Syntactically,
these rules are based on a broad subclass of safe domain relational calculus queries. We propose a new definition of support and confidence for entity-relationship rules and for the frequency of entity-relationship queries. We prove that the definition of frequency satisfies standard probability axioms and the Apriori property.
\end{abstract}

\section{Introduction}

One of the goals of data mining is to discover interesting relationships from
data. Association rules express relationships that hold with sufficient
frequency but not always. For example, it may be the case that not all
managers earn over \$60,000 a year, but that 90\% of managers do. The logical
form of an association rule is that of an implication $p\rightarrow q $ where
$p$ and $q$ hold together sufficiently often (the ``support" of the rule) and
$q$ holds sufficiently often given that $p$ holds (the ``confidence" of the
rule). The traditional 
concept of association rules severely limits the complexity
of the expressions $p$ and $q$ and thereby limits the class of 
relationships a data miner can capture. Essentially, $p$ and $q$ may be only
simple conjunctions, like an itemset. Thus we cannot have rules based on
Boolean combinations, such as negations or nested combinations. An example of
a relationship involving a negation would be a negative factor, such as
``students who have not taken an introductory database course do poorly in
datamining courses". An example of a nested Boolean combination would be ``students who are math majors or computer science majors, and who have done well in a discrete mathematics course or in an algorithms course, do well in complexity theory".
Another class of relationships that association rules
cannot express involves quantification and relating objects to each other. An
example would be the rule ``residents who have a neighbour with high incomes tend to have a high income themselves".

The goal of this paper is to extend the concept of an association rule to a
large class of expressions that we refer to as \emph{entity-relationship
queries }(ER queries). Intuitively, entity-relationship queries express
dependencies among entities and their properties. Entity-relationship queries
are a large subclass of the safe queries. Safe queries correspond to an
expressive subset of first-order logic that allows for nested Boolean
expressions and quantification. We provide a definition of the frequency %
of an entity-relationship query. This extends the
notion of an association rule to implications of the form $p\rightarrow q$
where $p \wedge q$ is an ER query; we refer to rules of this form as
\emph{entity-relationship rules}.
From our definition of the frequency of an ER
query we immediately obtain a definition of the support of an ER rule, namely
the frequency $fr(p\wedge q)$.

Our definition of frequency for ER queries
generalizes previous work on defining association rules in a multi-relational setting. \cite{negation-ref} discusses extending itemset rules with negations and motivates the usefulness of this extension. The query extension approach of the \warmr system \cite{bib:query-ex} presents a special class of entity-relationship rules that allows conjunctions of nonnegated statements and existential quantification. Our concept  of ER rules 
features in addition negations, universal quantification, nested quantifiers, and nested Boolean combinations. Thus one contribution of this paper is an extended rule format. A characteristic that distinguishes our approach from previous work is that previous approaches assume a given target table that defines a base set of tuples for evaluating the support of a query. In contrast, we start with a query and define a natural base set of tuples for evaluating the support of the query. We can think of this approach as dynamically generating entity sets for a given query rather than evaluating queries with respect to a fixed entity set. Thus the second main contribution of this paper is a new definition of support  for rules in our extended format.

The paper is organized as follows. First we review basic relational database concepts such as the relational schema and the domain relational calculus. Then we introduce the concept of an entity query and define the frequency of a query in this class of queries. This definition provides the basis for the notion of an entity-relationship rule and for defining the support of an entity-relationship rule. We compare entity-relationship queries to frequent itemsets and to the rule language of the \warmr system.
The final section establishes sevveral important formal properties of query frequencies as we define them and shows that they satisfy the Apriori property, that is, the frequency of a conjunction is no greater than the frequency of its conjuncts.


\section{Entities in the Domain Relational Calculus}
This section presents standard background material from database theory. The first subsection reviews relational schemas, and introduces the new concept of an \emph{entity field}. Semantically, entity fields are those that store values (constants) that refer to entities. The second subsection defines the standard notion of a safe query in the domain relational calculus, and the third introduces a subclass of safe queries that we term {\em entity-relationship queries}.

\subsection{Entities in Relational Schemas}

We begin with a standard \textbf{relational schema} containing a set of tables, each with key fields, 
descriptive attributes, and possibly foreign key pointers. We use the notation $\T$ to refer to a generic table that may represent either an entity set or a relationship set, and for an index we use $\T^i$. A field named $\name$ in table $\T$ is denoted by $\T.\name$. Table \ref{table:tv-schema} shows a relational schema for a TV survey database; this example is adapted from
\cite[Sec.2]{bib:turks}. Tables \ref{table:WeekdayTV}--\ref{table:areas} display {\em relation instances} for the TV survey schema.

\begin{table}[tbp] \centering
\begin{tabular}
[c]{|l|}\hline
TV-Program(\underline{Prog-Name}:string)\\
TV-Station(\underline{Station-Name}:string, Area:integer)\\
WeekdayTV\textbf{(}\underline{TV-Program:string,TV-Station:string}%
,Viewers:integer,Sponsor:string)\\
WeekendTV\textbf{(}\underline{TV-Program:string,TV-Station:string}%
,Viewers:integer,Sponsor:string)\\\hline
\end{tabular}
\caption{A relational schema for a TV survey model. Key fields are underlined. The schema lists TV programs and stations, and records for each combination of weekday program and station, how many viewers view the program on that station, and who sponsors the program. The same information is recorded for weekend programs.}
\label{table:tv-schema}
\end{table}%

\begin{table}[tbp] \centering
\begin{tabular}
[c]{|l|l|l|l|}\hline
\underline{TV-Program} & \underline{TV-Station} & Viewers & Sponsor\\\hline
Gilmore & Global & 10 & Avon\\\hline
Gilmore & CBS & 12 & La Senza\\\hline
Hockey Night & CBC & 20 & RBC\\\hline
\end{tabular}
\caption{Television Survey: Weekday TV.}\label{table:WeekdayTV}%
\end{table}%
%

\begin{table}[tbp] \centering
\begin{tabular}
[c]{|l|l|l|l|}\hline
\underline{TV-Program} & \underline{TV-Station} & Viewers & Sponsor\\\hline
Gilmore & Global & 8 & Avon\\\hline
Hockey Night & CBC & 14 & Schwab\\\hline
Simpsons & CBS & 10 & RBC\\\hline
Daily Show & CBC & 6 & La Senza\\\hline
\end{tabular}
\caption{Television Survey: Weekend TV.}\label{table:WeekendTV}%
\end{table}%
%

\begin{table}[tbp] \centering
\begin{tabular}
[c]{|l|c|}\hline
\underline{Station-Name} & Area\\\hline
Global & 1\\\hline
CBS & 2\\\hline
CBC & 3\\\hline
\end{tabular}
\caption{Television Survey: Stations and Areas.}\label{table:areas}%
\end{table}%

We assume that the tables in the relational schema can be divided into {\em entity tables} and {\em relationship tables.} This is the case whenever a relational schema is derived from an entity-relationship model (ER model) \cite[Ch.2.2]{bib:ullman}. Intuitively, an entity table corresponds to a type of entity, and a relationship table represents a relation between entity types. In our TV survey example, there are two types of entities: TV programs represented in the TV-Program table, and TV stations represented in the TV-Station table. We now introduce two assumptions concerning the relational schema that facilitate the definition of entity-relationship queries and their frequencies.

\begin{description}
\item[Unary Key Assumption] We assume that every entity table has a single key field.
\end{description}

The advantage of the unary key assumption is that given this assumption, a
single key field in the relational schema refers to a single entity.
The assumption holds in our TV survey schema because the two entity tables have key fields TV-Program.Prog-Name and TV-Station.Station-Name respectively. Although it is not always natural to define entities with a single key
field, there is no loss of generality because we can always form a single composite key field from a list of key fields. For example, if in a Professor table there are two key fields FirstName, LastName, we can form a composite key field $\langle\FirstName,\LastName\rangle$. Our second assumption is the following.

\begin{description}
\item[Global Name Assumption] We assume that for every entity $\e$, there is a
unique constant $\c$ such that in every table, the constant $\c$
denotes entity $\e$.
\end{description}

The global name assumption is important because it allows us to recognize when
the same entity occurs in different tables. In the AI literature, a similar assumption is often referred to as the ``unique name assumption" \cite[Ch.14]{bib:russell-norvig}.
The assumption
does not amount to a loss of generality because if the same constant $\c$ is
used in different tables to refer to different entities, we can simply index
$\c$ to distinguish these occurrences. For example, if we have two different transaction tables Transaction1 and Transaction 2, and there is a transaction 1
in both, we could change
the entry in the first table to refer to 1-1 and in the second table to refer to 1-2. A
natural alternative to indexing constants would be to adopt a convention to
the effect that a key field $\T.key$ in table $\T$ refers to different entities
than key field $\T^{\prime}.key$ in table $\T^{\prime}$ if and only if the names
of the key fields in the two tables are different. For example, if we have a
table for Employees and another for Managers, labelling the key field in each
table as ``ssn" indicates that a given social security number refers to the
same person no matter where it appears. In contrast, labelling the key field
in the Transactions1 table ``T1-number" and the key field in the
Transactions2 table as ``T2-number" indicates that the transaction
numbers in different tables refer to different transactions.

In many applications, the global name assumption is enforced through foreign key constraints. To illustrate, in the TV example, we may suppose that the field WeekdayTV.TV-Station is a foreign key pointer to the field TV-Station.Station-Name, and that the field WeekendTV.TV-Station is a foreign key pointer to the same field. So the string constant ``CBS" refers to the CBS
network represented in the TV-Station table, whether ``CBS" appears in an instance of the WeekdayTV relation or in an instance of the WeekendTV relation.

Given the unary key and global name assumptions, the following is a valid definition of how tables, key fields and constants are associated with entities.

\begin{definition}
Let $\D$ be a database instance.

\begin{enumerate}
\item An \textbf{entity table} is a table $\T$ with a single key field.

\item An field is an \textbf{entity field} if (1) the field is the key of an entity table, or (2) the field is a foreign pointer to the key of an entity table.

\item A constant $\c$ is an \textbf{entity constant} if $\c$ appears in an entity field.
%
\end{enumerate}
\end{definition}

\emph{Examples}. Let $\D$ be the TV survey database instance from
Tables \ref{table:WeekdayTV}--\ref{table:areas}. The entity keys
are TV-Program.Prog-Name, TV-Station.Station-Name,
WeekdayTV.TV-Program, WeekdayTV.TV-Station, Week-endTV.TV-Program,
WeekedTV.TV-Station. Entity constants include ``CBS" and
``Simpsons".

Next we review the domain relational calculus, which is a logical query language based on a given relational schema.

\subsection{Safe Queries in the Domain Relational Calculus}

We first define the formal language of the domain relational calculus, including the well-formed formulas of the calculus. Then we define an important subclass of formulas known as safe queries. Our presentation follows the standard approach, see for example \cite[Ch.3]{bib:ullman}.

\subsubsection{The Formal Language of the Domain Relational Calculus}
In the domain relational
calculus (DRC), for every table $\T^i$ in the database schema there is exactly one predicate $\P_i$ in the
logical language. The number of fields in the table $\T^i$ is the arity of the predicate $\P_i$. If $\T^i$ is an entity table, then $\P_i$ is an \textbf{entity predicate}. By the unary key assumption, an entity table $\T^i$ has a single key field; we adopt the convention that the key field is the first argument in the entity predicate $\P_i$.
The complete logical
vocabulary of the DRC is listed in Table \ref{table:vocab}.

\begin{table}[tbp] \centering
\begin{tabular}
[c]{|l|l|l|}\hline
Symbol Type & Notation & Comment\\\hline
Constants & $c_{1},c_{2},...$ & At most countably many constants \\\hline
Predicate Symbols & $P_{1},P_{2},..,P_{k}$ & Exactly one predicate for each
table $T^{i}$\\\hline
Logical Symbols & $\exists,\forall,\wedge,\vee,\lnot$ & \\\hline
Comparison Operators & $=,<,>,\leq,\geq,\neq$ & \\\hline
\end{tabular}
\caption{The Basic Vocabulary of our DRC language for a given database schema $\D$ with tables
$T^1,...,T^k$.} \label{table:vocab}%
\end{table}%

\emph{Example}. In the TV survey model, we have the predicates 
shown in Table \ref{table:example-vocab}.
\begin{table}[tbp] \centering
\caption{Predicates of our Logical Query Language for the TV survey
model.}\label{table:example-vocab}%
\begin{tabular}
[c]{|l|c|}\hline
\emph{Predicates} & Arity\\\hline
TV-Program(PN) & 1\\\hline
TV-Station(SN,A) & 2\\\hline
WeekdayTV(PN,SN,V,S) & 4\\\hline
WeekendTV(PN,SN,V,S) & 4\\\hline
\end{tabular}
\end{table}%

Thus we may write $WeekdayTV(\mbox{``GilmoreGirls",``CBS"},12, \mbox{``La Senza"})$ to assert that ``Gilmore
Girls" is shown on ``CBS" on weekdays, with 12,000 viewers, and sponsored by La Senza.
The notion of a well-formed formula is the usual one for this vocabulary.

\begin{definition} \label{def:safe-query}
Well-Formed Formulas of the Domain Relational Calculus

\begin{enumerate}
\item A constant $c$ or variable $X$ is a term.


\item If $P$ is a predicate symbol of arity $k$ and $t_{1},..,t_{k}$ are $k$
terms, then $P(t_{1},...,t_{k})$ is an atomic formula.

\item If $t,t^{\prime}$ are two terms, then a 
comparison
$t_{1}\theta t_{2}$ is an atomic formula.

\item If $F$ is a formula and $X$ is a variable, then $\lnot F,\exists
X.F,\forall X.F$ are formulas.

\item If $F_{1}$ and $F_{2}$ are formulas, then so are $F_{1}\wedge F_{2}$ and
$F_{1}\vee F_{2}$.

\item All formulas are formed by the repeated application of the previous rules.
\end{enumerate}
\end{definition}

\emph{Examples. }Table \ref{table:example-express} gives examples of valid
expressions and their types pertaining to the TV survey.\emph{\ }%
\begin{table}[tbp] \centering
\caption{Examples of Valid Expressions for the database schema for the TV survey.}\label{table:example-express}%
\begin{tabular}
[c]{|l|l|}\hline
Expression & Type\\\hline
$V \geq10$ & atomic formula with $V$ free\\\hline
$\exists S.\exists SN.\exists V.WeekdayTV(P,SN,V,S)$ & quantified formula with $P$ free\\\hline
$\exists S.\exists SN.\exists V.WeekdayTV(P,SN,V,S) \wedge V \geq10\wedge$ & conjunction of\\
$\exists S.\exists SN.\exists V.WeekendTV(P,SN,V,S) \wedge V \geq10$ & quantified
formulas\\\hline
\end{tabular}
\end{table}%

We next define the result or output of a DRC query.
The first step is to define what ground formulas are satisfied in a database
instance $\D$; a formula is \textbf{ground} if it contains no
variables.
The second step is to define which closed queries $F$ with no free variables
are satisfied in a database instance $\D$; as usual in logic, we
write $\mathcal{D\models}F$. Let $F[X_{1}/t_{1},..,X_{k}/t_{k}]$ be the
formula that results from replacing all free occurrences of each $X_{i}$ in
$F$ with the term $t_{i}$.

\begin{enumerate}
\item If $t,t^{\prime}$ are two constants, then $\D\models t\theta
t^{\prime}$ iff $t \theta t^{\prime}$ holds.

\item $\D\models P_{i}(c_{1},..,c_{k})$ iff $\langle c_{1},..,c_{k} \rangle$ is a tuple in table $T^{i}$.

\item $\D\models F_{1}\vee F_{2}$ iff $\D\models F_{1}$ or
$\D\models F_{2}$; similarly $\D\models F_{1}\wedge F_{2}$
iff $\D\models F_{1}$ and $\D\models F_{2}$; and
$\D\models\lnot F_{1}$ iff $\D\nvDash F_{1}$.

\item $\D\models\exists X.F$ iff there is a constant $c$ in the DRC language such that
$\D\models F[X/c]$; similarly $\D\models\forall X.F$ iff for
all constants $c$ we have $\D\models F[X/c]$.
\end{enumerate}

Let $F(X_{1},..,X_{m})$ be a query with free variables $X_{1},..,X_{m}$. Then
on database instance $\D$ the query $F$ returns the set of all tuples
that make $F$ true when substituted in $F$. Formally, we write%

\[
\tuples(\F)\equiv\{\langle c_{1},...,c_{m} \rangle:\D\models
F[X_{1}/c_{1},..X_{m}/c_{m}]\}.
\]

This definition assumes that the constants in the language include all constants that appear in the database tables, which involves no loss of generality.

\emph{Examples.} Let $\D$ be the database instance from Tables
\ref{table:WeekdayTV}--\ref{table:areas}. Table \ref{table:example-results}
shows the results of our example queries for this database instance.
\begin{table}[tbp] \centering
\begin{tabular}
[c]{|l|l|}\hline
Query Formula $F$ & Result $\tuples(\F)$\\\hline
$F_{1}=\exists S.\exists SN.\exists V.WeekdayTV(P,SN,V,S) \wedge V \geq10$ & \{``Gilmore",``Hockey
Night"\}\\\hline
$F_{2}=\exists S.\exists SN.\exists V.WeekendTV(P,SN,V,S) \wedge V \geq10$ & \{``Hockey
Night",``Simpsons"\}\\\hline
$F_{1}\wedge F_{2}$ & \{``Hockey Night"\}\\\hline
\end{tabular}
\caption{Results of Query Formulas on the database instance $\D$
from Tables \ref{table:WeekdayTV}--\ref{table:areas}.}\label{table:example-results}%
\end{table}%

\subsubsection{Safe Queries}
\label{sec:safe-queries}

It is customary to restrict the set of formulas that may serve as free variables in a query (``query
variables" for short) to ensure that the result set of tuples satisfying query formulas are
bounded and ``domain-independent" \cite[Ch.3.8]{bib:ullman}. To this end we adopt the notion of a safe query. The intuition behind this concept is that the results of safe queries should be restricted to selection conditions applied to (combinations of) tables in the database. For example, the query $\lnot TVProgram(\X)$ with free variable $\X$ is not safe because the range of constants satisfying this query is not bound by any  table in the database.
The key idea in the definition of safe query is to conjoin a query formula $\F$ to a restriction of the form $\P \wedge \F$ where $\P$ is a basic predicate in the language and hence refers to a table in the database. As is well-known, the expressive power of safe queries is exactly equivalent to that of relational algebra \cite{bib:codd}.
Safe queries are formally defined as follows \cite[Ch.3.8]{bib:ullman}.

\begin{enumerate}
\item Replace the $\forall X$ quantifier by $\lnot\exists X\lnot$.

\item Whenever $\vee$ is used to connect $F_{1}\vee F_{2}$, the two formulas
have the same set of free variables.

\item
Consider any maximal subformula consisting of the conjunction of
one or more formulas $F_{1}\wedge...\wedge F_{m}$. Then all variables $X$
appearing free in any of the $F_{i}$ must be limited as follows. The variable
$X$ must be free in some \emph{non-negated} $F_{i}$ satisfying \emph{one} of the
following conditions.

\begin{enumerate}
\item $F_{i}$ is not a comparison.

\item $F_{i}$ is $X = c$ 
where $c$ is a constant. \label{clause:key-field}

\item $F_{i}$ is $X=Y$, 
and $Y$ is limited.
\end{enumerate}

\item A $\lnot$ operator may apply only to a formula in a conjunction of the
type discussed in the previous rule.
\end{enumerate}

\emph{Examples}. Table \ref{table:safe-queries} gives examples of
safe and unsafe queries for the TV database schema from Table
\ref{table:tv-schema}.
\begin{table}[tbp] \centering
\begin{tabular}
[c]{|l|l|}\hline
Query Formula $F$ & Safe?\\\hline
$F_{1}=\exists S.\exists SN.\exists V.WeekdayTV(P,SN,V,S) \wedge V \geq10$ & yes\\\hline
$F_{2}=\exists S.\exists SN.\exists V.WeekendTV(P,SN,V,S) \wedge V \geq10$ &
yes\\\hline
$F_{1}\wedge F_{2}$ & yes\\\hline
$F_{1}\vee F_{2}$ & yes\\\hline
$\lnot F_{1}$ & no\\\hline
$\lnot F_{1}\vee F_{2}$ & no\\\hline
$F_{1}\wedge\lnot F_{2}$ & yes\\\hline
\end{tabular}
\caption{Examples of safe and unsafe queries for the TV survey database
schema of Table ~\ref{table:tv-schema}.}\label{table:safe-queries}%
\end{table}%

This completes our review of basic concepts from relational database theory.
We now come to the restriction of safe queries to
entity-relationship queries.

\subsection{Definition of Entity-Relationship Queries}

The basic idea behind our definition of an ER query is that free variables
should be limited in such a way as to guarantee that they must refer to
entities. 
Intuitively, an ER query is one whose free variables refer to
entities. The precise definition is as follows.

\begin{definition}
Let $\D$ be a database instance.\label{def:entity-var}

\begin{enumerate}
\item A variable $X$ is an \textbf{entity variable candidate} for a DRC formula $F$ if
\begin{enumerate}
\item $X$ is not quantified over in any part of $F$
\item if an expression $X \theta t$ appears in $F$, then $\theta$ is = or $\neq$, and if $t$ is a constant $c$, then $c$ is an entity constant in $\D$.
\item if an expression $P(\_,X,\_)$ appears in $F$, then the argument position of $X$ in $P(\_,X,\_)$ is an entity field.
\end{enumerate}
\item A variable $X$ is an \textbf{entity variable} for $F$ if
\begin{enumerate}
\item $X$ is an entity variable candidate for $F$, and
\item if an expression $X = Y$ or $X \neq Y$ appears in $F$, then $Y$ is an entity variable candidate.
\end{enumerate}

\item An \textbf{entity-relationship (ER) query} $F$ for database instance
$\D$ is a safe DRC query such that all the free variables in $F$ are
entity variables for $F$ given $\D$.
\end{enumerate}
\end{definition}

\emph{Examples. }Let $\D$ be the TV survey database instance from
Tables \ref{table:WeekdayTV}--\ref{table:areas}. In the formula $$\exists
S.\exists SN.\exists V.WeekdayTV(P,SN,V,S) \wedge V \geq10$$ the variable $P$ is an entity variable, and
so the formula is an ER query. The formula $$\exists S.\exists SN.WeekdayTV(\mbox{``Gilmore Girls"},SN,V,S)$$ is safe but not an ER query because the free variable $V$ is not an entity variable.

\section{The Frequency of Entity-Relationship Queries}

Our basic idea is that the limiting conditions in safe queries specify the
domain from which values for a free variable $\X$ are to be drawn. Once the domain for the free variables is defined for a given formula $\F$, we can take the frequency of the formula $\F$ to be the number of assignments to the free variables that satisfy the formula divided by the size of the domain for the formula. Safe queries are a natural class of queries for this approach because these queries specify the range from which result tuples may be drawn by restricting these results to subsets of tables in the database (cf. Section \ref{sec:safe-queries}).

The main
issue in our definition concerns the correct domain for conjunctions or intersections. For a
simple example, consider a database schema with two entity tables Professor and Customer. The query $Professor(X)\wedge Customer(X)$ returns
entities that are both professors and customers.
 What should be the base domain for this
query? If there are many more customers than professors, we may get quite
different frequency counts if we take the base domain to be Professor 
than if
we take it to be Customer. 
So neither of these seems the right choice.
Intuitively the base domain should be a symmetric function of the
two classes mentioned in the query. The two natural symmetric set-theoretic
operations are intersection and union. If we take the intersection as the base
domain, the frequency of conjunctions without further selection conditions is
always 100\%, which does not seem right. In particular for our ultimate goal
of defining the support of association rules, this is unsatisfactory. Our
proposal is therefore to use the \emph{union} of the two entity sets involved
in the conjunction. Another way to look at the union is that it represents a
kind of closed world assumption: If Professors and Customers are the only entity types
mentioned in the selection conditions of the query, then the members of these entity types are
exactly the potential answers to the query.

The closed world assumption is also the basis for our frequency definition for queries with negation. For example, consider a safe query such as $Professor(\X) \wedge \lnot Customer(\X)$. Since Professors and Customers are the only entity types mentioned in this query, we take the base domain again to be the union of these two sets. The fact that Professors are mentioned positively and Customers negatively does not make a difference to the base domain, but it does make a difference to the result of the query and hence to its frequency.


On the basis of this proposal, we can now recursively assign a domain to an
entity variable in a formula $F$ given a database instance $\D$. We begin with just one free query variable and then tackle the more complicated case of queries with more than one free variable.

\subsection{Definition of Frequency for Queries With One Free Query Variable}

We denote the base domain of an entity variable $\X$ in a query $\F$ relative to a database instance $\D$ as $\dom(F,\X)$. As we think of variable $\X$ as
referring to the domain $\dom(F,\X)$, we term $\dom(F,\X)$ the \textbf{reference domain} of $\X$ in the context of query $F$.

\begin{definition} \label{def:ref-domain-one}
Let $\D$ be a database instance with ER formula $F$.

\begin{enumerate}
\item If $F$ is $P_{i}(t_{1},..,t_{k})$, and $\X$ occurs in $\F$, then $\dom(\F,\X)= \pi_{\X} [\tuples(\F)]$. If $\X$ is not a free variable in $\F$, then $\dom(\F,\X)= \emptyset$.
Here we think of $\tuples(\F)$ as a relation whose columns correspond to the free variables of $\F$. For example, the query $p(X,Y,Z)$ returns a relation with triples, and we can think of the first column as named $X$ and the second as named $Y$. The expression $\pi$ refers to the projection operator of relational algebra (with elimination of duplicates).

\item Let $F$ be a single atomic comparison of the form $Y \theta t$ where $t$ is either a variable or a constant. If $F$ is $\X = c$, then $\dom(F,X) = \{c\}$. Otherwise $\dom (\F,\X)=\emptyset$.
%

%



\item If $F$ is $\lnot G$ for some formula $G$, then $\dom(\F,\X)=\dom(\G,\X)$.

\item If $F$ is $F_{1}\vee F_{2}$ or $F_{1}\wedge F_{2}$, then
$\dom(\F,\X)=\dom(\F_{1},\X)\cup \dom%
(\F_{2},\X)$.

\item If $\F$ is $\exists \Y.G$, where $\Y \neq \X$, then $\dom(\F,\X)=\dom(G,\X)$. If $\F$ is $\exists \X.G$, then $\dom(\F,\X) = \emptyset$.
\end{enumerate}

\end{definition}

\emph{Examples. }Let $\D$ be the TV survey database instance from
Tables \ref{table:WeekdayTV}--\ref{table:areas}. Table
\ref{table:example-refsort} gives examples of reference domains for various ER
queries.

\setlength{\tabcolsep}{2 pt}
\begin{table}[tbp] \centering

\begin{tabular}[c]{|l|}
\hline

Query Formula $\F$, Reference Domain $\dom(\F,\X)$
\\

\hline

$F_{1}$=
$\exists S.\exists SN.\exists
V.WeekdayTV(P,SN,V,S) \wedge V \geq10$\\

$\dom(\F_1,\X)=$ 
$ \{$``Gilmore"$,$``Hockey Night"$\}$\\

\hline

$F_{2}=$
$\exists
S.\exists SN.\exists V.WeekendTV(P,SN,V,S) \wedge V \geq10$ \\

$\dom(\F_2,\X)=$
$\{$``Gilmore"$,$``Hockey
Night",``Simpsons",``Daily
Show"$\}$\\

\hline

$F_3 = $
$F_{1}\wedge F_{2}$ \\

$\dom(\F_3,\X)=
$
$\{$``Gilmore"$,$``Hockey
Night",``Simpsons",``Daily Show"$\}$\\

\hline

$F_4 = $
$F_{1}\vee F_{2}$ \\

$\dom(\F_4,\X)=$
$\{$``Gilmore"$,$``Hockey
Night",``Simpsons",``Daily
Show"$\}$\\

\hline

$F_5=$
$F_{1}\wedge\lnot F_{2}$ \\

$\dom(\F_5,\X)=$
$\{$``Gilmore"$,$``Hockey
Night",``Simpsons",``Daily Show"$\}$\\

\hline
\end{tabular}
\caption{Reference Domains for various formulas in the TV survey
database
instance $\D$ from Tables \ref{table:WeekdayTV}--~\ref{table:areas}.}\label{table:example-refsort}%

\end{table}

As this definition shows, we think of basic predicates as specifying the range
from which entities are drawn. Conditions of the form $p(t_{1},...,X,...t_{k}%
)$ or $X=c$ 
we view as ``direct bounds"
that determine the reference domain of $X$. Variable equations of the form $X=Y$ we
view as ``selection conditions" that are applied after an entity has been
specified.
These do not affect the reference domain of $X$ but only the result
of the query. 
Another type of selection are restrictions on descriptive attributes, such as $\V \geq 10$ in the queries in Table \ref{table:example-refsort}.

Now the frequency of an 
ER query is defined as follows.

\begin{definition}
Let $F$ be an ER query with free variable $\X$ such that $\dom(\F,\X) \neq \emptyset$. Then%

\[
fr_{\D}(F)\equiv \frac{|\tuples(\F)|}{|\dom(\F,\X)|}.
\]
\end{definition}

In Section \ref{sec:apriori} we establish several formal properties of the frequency of a query according to this definition, for example that the frequency is a number between 0 and 1.

\emph{Examples.} Let $\D$ be the TV survey database instance from
Tables \ref{table:WeekdayTV}--\ref{table:areas}. Table
\ref{table:example-frequency} illlustrates the frequencies of various
queries.
\setlength{\tabcolsep}{3pt}
\begin{table}[tbp] \centering
\begin{tabular}
[c]{|l|c|}\hline
Query Formula $F$ & Frequency $fr_{\D}(F)$ \\\hline
$F_{1}=\exists S.\exists SN.\exists V.WeekdayTV(P,SN,V,S) \wedge V \geq10$ & $1$ \\\hline
$F_{2}=\exists S.\exists SN.\exists V.WeekendTV(P,SN,V,S) \wedge V \geq10$ & $1/2$ \\\hline
$F_{1}\wedge F_{2}$ & $1/4$ \\\hline
$F_{1}\vee F_{2}$ & $3/4$ \\\hline
$F_{1}\wedge\lnot F_{2}$ & $1/4$ \\\hline
\end{tabular}
\caption{Frequencies for various formulas in the TV survey database instance
$\D$ from Tables \ref{table:WeekdayTV}--\ref{table:areas}.}\label{table:example-frequency}%
\end{table}%

\subsection{Definition of Frequency of Queries With More Than One Free Variable}

We assign a domain to every
tuple of entity variables in a formula $F$ given a database instance $\D$,
which we denote as $\dom(\F,\{X_1,...,X_m\})$. Our basic idea is to consider a result tuple $\langle \c_1,...,\c_m \rangle$ as denoting a {\em composite entity} formed by combining $m$ single entities. For example, consider the rule $\Neighbour(\X,\Y) \wedge (\exists \I. Income(\X,\I) \wedge \I > \$100,000) \rightarrow (\exists \I. Income(\Y,\I) \wedge \I > \$100,00)$. (The symbol $\rightarrow$ does not denote logical implication but defines an association rule; see Section \ref{sec:rules}.) This says that if $\X$ has an income over \$100,000, then it is likely that a neighbour $\Y$ of $\X$ also has an income of \$100,000.
The support of this rule is the frequency of the query $\Neighbour(\X,\Y) \wedge (\exists \I. Income(\X,\I) \wedge \I > \$100,000) \wedge (\exists \I. Income(\Y,\I) \wedge \I > \$100,00)$. This query has two free variables $\X$ and $\Y$. The reference domain comprises the entries in the $\Neighbour$ table, that is, the pairs $\langle \X,\Y \rangle$ in the table. Other examples of natural composite entities include relations like reservations or purchases. The idea of treating tuples in a relational table as composite ``individuals" is familiar in the propositionalization literature \cite{bib:prop, bib:deraedt-abstract} (for example, chemical molecules may be treated as single entities although molecules are composed of different elements that are also represented in the relational schema). Applying this idea requires a further constraint on ER queries: the free variables $\{\X_1,...,\X_m\}$ must be ``bound together" in a limiting condition rather than separately. For example, the query $\P(\X) \wedge \X=\Y$ is a safe ER query but the answer pairs $\langle x,y \rangle$ are not bound to the key fields of any tuple; an example of the same character is the query $\P(\X) \wedge \Q(\Y)$. 
To rule out such cases, we impose the following condition.

\begin{definition} A \textbf{literal} is an atomic formula or its negation.
An ER query $\F$ is \textbf{valid for variables} $\X_1,...,\X_m$ if for every maximal conjunction $\L = \L_1 \wedge ... \wedge \L_k$ consisting only of literals, $\L$ contains a conjunction of the form $\X_1 = \c_1 \wedge \cdots \wedge \X_m = \c_m$, or $\L$ contains a conjunct $\P(\t_1,...,\t_k)$ where all variables $\{\X_1,...,\X_m\}$ occur in $\P(\t_1,...,\t_k)$. An ER query $\F$ is \textbf{valid} if $\F$ is valid for the set of its free variables. \label{def:valid}
\end{definition}

Examples follow below in this section. In the case with only one free query variable $X$, the definition of safe query implies that every entity query is valid. Now let us consider the definition of a reference domain for valid ER queries with one or more free variables. As in the case with just one query variable,
we term $\dom(F,\{X_1,...,X_m\})$ the \textbf{reference domain} of $\{\X_1,...,\X_m\}$ in the context of query $F$.
Consider the basic case of an atomic formula $F = \P(t_1,...,t_m)$ first. In keeping with the idea behind safe queries, we can think of such formulas as specifying a basic range for the result tuples in a query. So suppose that the free variables in the atomic formula are $Y_1,Y_2,..,Y_k$. If our query variables $X_1,...,X_m$ are not {\em all} contained in the set $\{Y_1,Y_2,..,Y_k\}$, we consider that the ``composite key" $X_1,...,X_m$ does not appear in the query, and $\dom(F,\{X_1,...,X_m\})= \emptyset$. Otherwise we consider the query result $\tuples(\F)$ as a relation with $k$ columns, of which $m$ are named $X_1,..,X_m$. For example, the query $p(X,Y,Z)$ returns a relation with triples, and we can think of the first column as named $X$ and the second as named $Y$. Thus we can take $\pi_{\langle \X_1,...,\X_m \rangle}\tuples(\F)$ to be the reference domain of the entity variables $X_1,X_2,...,X_m$ in the query $F$. This leads to the following inductive definition. The main difference with the definition for a single query variable is that we need to treat conjunctions like $\X_1 = \c_1 \wedge \cdots \X_m = \c_m$ as a single compound statement.



\begin{definition} \label{def:ref-domain}
Let $\D$ be a database instance with ER formula $\F$ and let
$X_1,..,X_m$ be a list of variables.

\begin{enumerate}
\item If $F$ is $\P(t_{1},..,t_{k})$, and all variables
$X_1,..,X_m$ occur in $\P(t_{1},..,t_{k})$, then
$\dom(\F,\{X_1,...,X_m\})= \pi_{\langle X_1,...,X_m \rangle}
\tuples(\F)$, where $\pi$ is the projection operation of
relational algebra. Otherwise $\dom$ $(\F,\{X_1,...,X_m\})=
\emptyset$.
\item Let $F$ be a single atomic comparison of the form $Y \theta t$ where $t$ is either a variable or a constant.

\begin{enumerate}
\item Suppose that $m = 1, Y = X_1$ and the comparison is $X_1 = c$ (i.e., we just have a single free variable $X_1$ and the atomic formula requires $X_1$ to be equal to a constant $c$.) In that case $\dom(F,\{X_1\}) = \{c\}$.
\item Otherwise $\dom(F,\{X_1,...,X_m\}) = \emptyset$.
\end{enumerate}

\item Let $F$ be a maximal conjunction of $k >1$ formulas, such that $F = C_1 \wedge \cdots \wedge C_k$.

\begin{enumerate}
\item If $F$ is a conjunction of the form $C \wedge X_1 = c_1 \ldots \wedge X_m = c_m$, then $\dom(F,\{X_1,...,X_m\}) = \dom(C,\{X_1,...,X_m\}) \cup \{\langle c_1,...,c_m \rangle\}$. 
\item Otherwise $\dom(F,\{X_1,...,X_m\}) = \bigcup_{i=1}^{k} \dom(C_i,\{X_1,...,X_m\})$. 
\end{enumerate}

\item If $F$ is $F_{1}\vee F_{2}$, then
$\dom(F,\{X_1,...,X_m\})=\dom(F_{1},\{X_1,...,X_m\})\cup \dom%
(F_{2},\{X_1,...,X_m\})$. 

\item If $F$ is $\lnot G$ for some formula $G$, then $\dom(F,\{X_1,...,X_m\})=\dom(G,\{X_1,...,X_m\}$.

\item If $F$ is $\exists Y.G$, where $\Y \not\in \{X_1,...,X_m\}$, then $\dom(F,\{X_1,...,X_m\})=\dom(G,\{X_1,...,X_m\})$. If $\F$ is $\exists \X_i.G$ for some $\X_i \in \{X_1,...,X_m\}$, then $\dom(F,\{X_1,...,X_m\}) = \emptyset$.
\end{enumerate}

\end{definition}

It is easy to check that this definition agrees with Definition \ref{def:ref-domain-one} for queries with just one free variable.

{\em Examples.} Consider the query ``find all program-station pairs that achieve a viewership of over 10,000 on both weekdays and weekends". In the domain relational calculus, this query may be formulated as $[\exists S.\exists \V. WeekdayTV(P,SN,V,S) \wedge V > 10] \wedge [\exists S.\exists \V. WeekemdTV(P,SN,V,S) \wedge V > 10]$. Table \ref{table:ref-dom-multiple} shows the calculation of the reference domain for this formula on the database instance of Tables \ref{table:WeekdayTV}--\ref{table:areas}.

\setlength{\tabcolsep}{3pt}
\begin{table}[tbp] \centering
\begin{tabular}
[c]{|l |}\hline

Query Formula $F$, Reference Domain
$\dom(F,X)$\\
\hline $\F_{1}= \exists S.\exists \V. WeekdayTV(P,SN,V,S)
\wedge V > 10$ \\

$\dom(F_1,X)=\{\langle \mbox{\small ``Gilm.",``Glo."} \rangle,
\langle \mbox{\small``Gilm.",``CBS"}\rangle, \langle
\mbox{\small``Hock. N.",``CBC"} \rangle\}$\\\hline

$F_{2}= \exists S.\exists \V. WeekendTV(P,SN,V,S) \wedge V > 10$\\
$\dom(F_2,X)=\{\langle \mbox{\small``Gilm.",``Glo."} \rangle,
\langle \mbox{\small``Hock. N.", ``CBC"}\rangle,
\langle\mbox{\small``Simps.",``CBS"}\rangle, \langle
\mbox{\small``Daily Sh.",``CBC"} \rangle\}$\\\hline $F_3
=F_{1}\wedge F_{2}$\\ $\dom(F_3,X)=\{\langle
\mbox{\small``Gilm.",``Glo."} \rangle, \langle
\mbox{\small``Gilm.",``CBS"}\rangle, \langle \mbox{\small``Hock.
N.",``CBC"} \rangle, \langle \mbox{\small``Simps.",``CBS"}\rangle,
\langle \mbox{\small``Daily Sh.",``CBC"} \rangle\}$\\\hline
\end{tabular}
\caption{Reference Domains for various formulas in the TV survey database
instance $\D$ from Tables \ref{table:WeekdayTV}--\ref{table:areas}. The free variables query variables are $\P$ and $\SN$, corresponding to pairs of programs-stations.}\label{table:ref-dom-multiple}%
\end{table}%


Now the frequency of an ER query is defined as follows.

\begin{definition}
Let $F$ be an ER query whose free variables are $X_{1},..,X_{m}$ where $\dom(\{\X_1,...,\X_m\},\F) \neq \emptyset$. Then 

$$
fr_{\D}(F)\equiv\frac{|\tuples(\F)|}{|\dom(\F,\{\X_1,...,\X_m\})|}.
$$
\end{definition}
Table
\ref{table:example-frequency-multiple} illustrates the frequencies of various
queries.
\begin{table}[tbp] \centering
\begin{tabular}
[c]{|l|c|}\hline
Query Formula $F$ & Frequency $fr_{\D}(F)$\\\hline
$\F_{1}= \exists S.\exists \V. WeekdayTV(P,SN,V,S) \wedge V > 10$ & $2/3$\\\hline
$F_{2}= \exists S.\exists \V. WeekendTV(P,SN,V,S) \wedge V > 10$ & $1/4$\\\hline
$F_{1}\wedge F_{2}$ & $1/5$\\\hline
$F_{1}\vee F_{2}$ & $2/5$\\\hline
$F_{1}\wedge\lnot F_{2}$ & $1/5$\\\hline
\end{tabular}
\caption{Frequencies for various formulas in the TV survey database instance
$\D$ from Tables \ref{table:WeekdayTV}--\ref{table:areas}. The free variables query variables are $\P$ and $\SN$, corresponding to pairs of programs-stations.}\label{table:example-frequency-multiple}%
\end{table}%

\section{Entity-Relationship Rules}
\label{sec:rules}
We finally obtain the notion of an ER association rule, or ER rule for short.

\subsection{Definition of Confidence and Support for ER rules}

Given the concepts we have developed so far, the definition of confidence and support for an entity-relationship rule are straightforward.

\begin{definition} \label{def:rule}
Let $\D$ be a database instance.

\begin{enumerate}
\item An \textbf{ER association rule} is an implication of the form $F\rightarrow G$,
where the free variables of $G$ are the same as or contained in the free variables of $F$,
and $\F \wedge \G$ is a valid ER query.


\item The \textbf{confidence} of an ER association rule $F\rightarrow G$ is given by
$$\con_{\D}(F\rightarrow G)\equiv\frac{|\tuples(\F \wedge \G)|}{|\tuples(\F)|}.$$

\item The \textbf{support} of an ER association rule $F\rightarrow G$ is given by $$\supp_{\D}(F\rightarrow G)\equiv fr_{\D}(F\wedge G).$$
\end{enumerate}
\end{definition}

As usual with association rules, the  implication $\F \rightarrow \G$ does not indicate logical implication (whenever $\F$ is true, so is $\G$) but instead denotes a probabilistic relationship.

\emph{Example.} Let $\D$ be the TV survey database instance from
Tables \ref{table:WeekdayTV}--\ref{table:areas}. Let $F_{1}$ be the formula
$$\exists S.\exists SN.\exists \V. WeekdayTV(P,SN,V,S) \wedge V \geq 10$$ 
and let $F_{2}$ be the formula $$\exists S.\exists SN.\exists \V. WeekendTV(P,SN,V,S) \wedge V \geq 10.$$
Consider the rule
$F_{1}\rightarrow F_{2}$. The support of this rule is $fr_{\D}(F_{1}\wedge
F_{2})=1/4$ (see Table \ref{table:example-frequency}). The confidence is
\[
\frac{|\{\text{``Gilmore",``Hockey Night"}\}\cap\{\text{``Hockey Night",``Simpsons"}%
\}|}{|\{\text{``Gilmore",``Hockey Night"}\}|}=1/2.
\]

Definition \ref{def:rule} completes our goal of providing a definition of confidence and support for general entity-relationship queries.

\subsection{Comparison With Other Rule Languages} \label{sec:comparison}

This section gives a brief comparison of our rule language and frequency definition to related rule languages. It is easy to see that the classic association rule approach based on frequent itemsets is a special case. For example, suppose we have two entity tables: Transactions(\underline{number}) that stores transactions, and Item(\underline{name}) for items, and a relational table TransItems(\underline{TransNumber},\underline{ItemName}) that indicates which items appear in which transactions. Then for a given item, say ``cola", the query $Transactions(X) \wedge TransItems(X,\mbox{``cola"})$ returns the set of transactions involving ``cola", and the frequency of this query is the frequency of these transactions among all transactions.

Antonie and Za\"{i}ane \cite{negation-ref} extend itemset rules with negations, and survey a number of search algorithms for finding frequent itemsets with negative conditions. Their search procedure is based on correlation analysis.

The \warmr system \cite{bib:query-ex} considers queries that are conjunctions of literals (e.g., $P(X,Y)$). The user specifies a target table $T$; the free query variables in a \warmr query are then bound to the key fields of $T$. If WeekdayTV is our target table, we would have two free query variables $P$ for program and $SN$ for station. All other variables are implicitly existentially quantified. For example, if Customer is the target table, the \warmr formula $Customer(A) \wedge Child(A,C) \wedge Buys(C,\mbox{``cola"})$ translates into the domain relational calculus as $\exists C. Customer(A) \wedge Child(A,C) \wedge Buys(C,\mbox{``cola"})$. If we assume that one of the conjuncts in a \warmr clause corresponds to the target table (e.g., $Customer(A)$), and all other appearances of the query variables are related to the target table by foreign key constraints (e.g., the first field in the Child table is a foreign key to the Customer table), then the reference domain as we have defined it is exactly the target table, and the frequency that \warmr assigns to a conjunction agrees with our definition. In this sense our definition of support for ER rules generalizes that for \warmr rules.



\section{The Probability Axioms and A Priori Property}
\label{sec:apriori}

In order to ensure that Definition \ref{def:ref-domain} yields well-defined probabilities, we verify three facts: (1) the frequency as defined never involves division by 0, so the frequency is well-defined. (2) The definition entails that frequencies are between 0 and 1 (inclusively). (3) The frequency of two mutually exclusive queries is the sum of their respective frequencies. This third property holds only with certain qualifications due to the restrictions on safe queries. The usual probability axioms include the requirement that (4) the probability of the whole space, or the ``certain event" is 1. We discuss the extent to which this property holds for our definition of frequency. Finally, we show the \apriori property: frequencies of conjunctions decrease monotonically, which is important for lattice search methods.

For the first fact, we have the following result. The notion of a valid ER query was specified in Definition \ref{def:valid}.



\begin{proposition}
Let $\F$ be a valid ER query whose free variables are $\X_1,...,\X_m$. 
Let $\D$ be any database instance (without empty tables). Then $\dom(\F,\{X_1,...,X_m\}) \neq \emptyset$.
\end{proposition}

\begin{proof}
If $\F$ is valid, then for every maximal conjunction $\L$ of literals that occurs in $\F$, we have $\dom(\L,\{X_1,...,X_m\}) \neq \emptyset$. Since the reference domains of more complex formulas are the union of the domains of their subformulas, it follows that  $\dom(F,\{X_1,...,X_m\}) \neq \emptyset$.
\end{proof}

The next proposition guarantees that the ratios assigned by Definition \ref{def:ref-domain} are properly bounded between 0 and 1.

\begin{proposition}
Let $\F$ be an ER query in which the variables $\X_1,\ldots,\X_m$ are free such that $\F$ is valid for these variables. Let $\D$ be a database instance. 
Then $\pi_{\langle \X_1,...,\X_m \rangle} \tuples(\F) \subseteq \dom(\F,\{\X_1,\ldots,\X_m\})$, where $\pi$ is the projection operation of relational algebra.
\end{proposition}

In the case in which $\X_1,\ldots,\X_m$ are exactly the free variables of $\F$, we have $\pi_{\langle \X_1,...,\X_m \rangle} \tuples(\F) = \tuples(\F)$, so the proposition implies that the ratio $\frac{|\tuples(\F)|}{|\dom(\F,\{\X_1,...,\X_m\})|}$ is between 0 and 1.

\begin{proof}
The proof is by induction on the structure of ER formula $\F$. We begin by noting two basic facts about valid formulas, which follow easily from Definitions \ref{def:safe-query}, \ref{def:valid}, and \ref{def:ref-domain}.

\begin{enumerate}
\item  If $\C = \C_1 \wedge ... \wedge \C_k$ is a maximal conjunction in $\F$, then $\C$ contains a conjunction $\X_1 = c_1 \ldots \wedge \X_m = c_m$ or a conjunct $\C_i$ that is a valid ER query. \label{clause:conjunction}
\item If $\F_1 \vee \F_2$ is a disjunction in $\F$, then both of the disjuncts are valid ER queries. \label{clause:disjunct}
\end{enumerate}


\begin{itemize}
\item If $\F$ is an atomic formula of the form $\P(t_{1},..,t_{k})$, then since $\F$ is valid for $\X_1,...,\X_m$, we have $\dom(F,\{\X_1,...,\X_m\})= \pi_{\langle X_1,...,X_m \rangle} \tuples(\F)$.

\item Let $F$ be a single atomic comparison of the form $Y \theta t$ where $t$ is either a variable or a constant. Since $\F$ is valid, it must be of the form $X_1 = c$  where $m=1$ (i.e., we just have a single free variable $X_1$ and the atomic formula requires $X_1$ to be equal to a constant $c$). So $\dom(F,\{X_1\}) = \{c\}$, and clearly $\pi_{\X_1}  \tuples(\F) \subseteq \{c\}$.

\item Let $F$ be a maximal conjunction of $k>1$ formulas, such that $F = C_1 \wedge \cdots \wedge C_k$.

\begin{enumerate}
\item If $\F$ is a conjunction of the form $\C \wedge \X_1 = c_1 \ldots \wedge \X_m = c_m$, then $\dom(F,\{X_1,...,X_m\}) = \dom(C,\{X_1,...,X_m\}) \cup \{\langle c_1,...,c_m \rangle\}$. Clearly $\pi_{\langle \X_1,...,\X_m \rangle} \tuples(\F) \subseteq \{\langle c_1,...,c_m \rangle\}$, which is a subset of  $\dom(F,\{X_1,...,X_m\})$.
\item Otherwise $\dom(F,\{X_1,...,X_m\}) = \bigcup_{i=1}^{k} \dom(C_i,\{X_1,...,X_m\})$. Since $\F$ is valid, by Observation \ref{clause:conjunction}
at least one of the conjuncts $\C_i$ is valid.
So by inductive hypothesis, $$\pi_{\langle \X_1,...,\X_m \rangle} \tuples(\C_i) \subseteq \dom(\{\C_i, \{\X_1,\ldots,\X_m\}).$$
Now since $\F$ is a conjunction involving $\C_i$, it follows that  $$\pi_{\langle \X_1,...,\X_m \rangle} \tuples(\F)\subseteq \pi_{\langle \X_1,...,\X_m \rangle} \tuples(\C_i)$$ and that $$\dom(\C_i,\{\X_1,\ldots,\X_m\}) \subseteq \dom(\F,\{\X_1,\ldots,\X_m\}),$$ which establishes the inductive hypothesis for this case.
\end{enumerate}

\item If $F$ is $F_{1}\vee F_{2}$, then by Clause \ref{clause:disjunct} of the definition of a safe query, both $\F_1$ and $\F_2$ are valid and contain all the variables $\{X_1,...,X_m\})$ as free variables. So $$\pi_{\langle \X_1,...,\X_m \rangle} \tuples(\F) = \pi_{\langle \X_1,...,\X_m \rangle} \tuples(\F_1) \cup \pi_{\langle \X_1,...,\X_m \rangle} \tuples(\F_2).$$ Also, by inductive hypothesis,
 $$\pi_{\langle \X_1,...,\X_m \rangle} \tuples(\F_1) \subseteq \dom(\F_1,\{\X_1,\ldots,\X_m\})$$ and $$\pi_{\langle \X_1,...,\X_m \rangle} \tuples(\F_2) \subseteq \dom(\F_2,\{\X_1,\ldots,\X_m\}),$$ and by definition $$\dom(\F,\{\X_1,\ldots,\X_m\}) = \dom(\F_1,\{\X_1,\ldots,\X_m\}) \cup \dom(\F_2,\{\X_1,\ldots,\X_m\}).$$ So
 $$\pi_{\langle \X_1,...,\X_m \rangle} \tuples(\F) \subseteq \dom(\F,\{\X_1,\ldots,\X_m\})$$ as required.

\item If $F$ is $\lnot G$ for some formula $G$, then $\F$ is not a safe query, hence not an ER query, and the claim holds vacuously.

\item If $F$ is $\exists Y.G$, then $\Y \not\in \{X_1,...,X_m\}$, since the variables $X_1,...,X_m$ are free in $\F$. So $$\dom(F,\{X_1,...,X_m\})=\dom(G,\{X_1,...,X_m\}),$$ and $$\pi_{\langle \X_1,...,\X_m \rangle} \tuples(\F) = \pi_{\langle \X_1,...,\X_m \rangle} \tuples(\G)$$ by the semantics of the existential quantifier. Clearly if $\F$ is valid, then so is $\G$, so by inductive hypothesis $$\pi_{\langle \X_1,...,\X_m \rangle} \tuples(\G) \subseteq \dom(\G,\{\X_1,\ldots,\X_m\})$$ which completes the inductive proof.

\end{itemize} \end{proof}





The third fundamental property of probabilities is {\em finite additivity}, that the frequency of two mutually exclusive events is the sum of the individual frequencies. The difficulty with this property is not that it fails for our frequency definition, but that it is not straightforwardly expressed in our language of safe queries. For example, a natural formulation of finite additivity would be to require that $fr_{\D}(\F) + fr_{\D}(\neg \F) =  fr_{\D}(\F \vee \neg \F)$. But if $\F$ is a safe query, then $\neg F$ is not safe, so the frequency $fr_{\D}(\neg F)$ is not defined. Another way to see the difficulty is to note that in standard probability theory (with a Boolean algebra of events), finite additivity is equivalent to the requirement that $Pr(A) = 1- Pr(\bar{A})$, where $\bar{A}$ is the complement of event $A$. But this cannot be expressed as a requirement on safe queries since the negation of a safe query is not itself safe. 

However, we can show a qualified version of finite additivity. If $\S$ and $\F$ are valid safe queries with the same free variables, then the formulas $\S \wedge \F$ and $\S \wedge \neg F$ are also valid safe queries. For these formulas we can show the following result.

\begin{proposition}
Let $\S$ and $\F$ be valid safe queries with the same free variables $\{X_1,...,X_m\}$. Then for any database instance $\D$ we have

$$fr_{\D}([\S \wedge \F] \vee [\S \wedge \neg \F]) = fr_{\D}(\S \wedge \F) + fr_{\D}(\S \wedge \neg \F) = \frac{\tuples(\S)}{\dom(\S,\{X_1,...,X_m\}) \cup \dom(\F,\{X_1,...,X_m\})}.$$
\end{proposition} 

\begin{proof}
This follows from the definitions: We have $\dom([\S \wedge \F] \vee [\S \wedge \neg \F], \{X_1,...,X_m\}) = \dom([\S \wedge \F],\{X_1,...,X_m\}) \cup \dom([\S \wedge \neg \F],\{X_1,...,X_m\})$, and since $\dom([\S \wedge \F],\{X_1,...,X_m\}) = \dom([\S \wedge \neg \F],\{X_1,...,X_m\}) = \dom(\S,\{X_1,...,X_m\}) \cup \dom(F,\{X_1,...,X_m\})$, it follows that 

$$\dom([\S \wedge \F] \vee [\S \wedge \neg \F],\{X_1,...,X_m\})= \dom(S,\{X_1,...,X_m\}) \cup \dom(F,\{X_1,...,X_m\}).$$

Clearly $\tuples([\S \wedge \F] \vee [\S \wedge \neg \F]) = \tuples(S)$, so 

$$fr_{\D}([\S \wedge \F] \vee [\S \wedge \neg \F]) = \frac{\tuples(\S)}{\dom(\S,\{X_1,...,X_m\}) \cup \dom(\F,\{X_1,...,X_m\})}.$$

Also, $fr_{\D}(\S \wedge \F) = \frac{\tuples(\S \wedge \F)}{\dom(\S,\{X_1,...,X_m\}) \cup \dom(\F,\{X_1,...,X_m\})}$ and $fr_{\D}(\S \wedge \neg \F) = \frac{\tuples(\S \wedge \neg \F)}{\dom(\S,\{X_1,...,X_m\}) \cup \dom(\F,\{X_1,...,X_m\})}$, so 
$$fr_{\D}(\S \wedge \F) + fr_{\D}(\S \wedge \neg \F) = \frac{\tuples(\S)}{\dom(\S,\{X_1,...,X_m\}) \cup \dom(\F,\{X_1,...,X_m\})},$$ which was to be shown.
\end{proof}

This result illustrates that two logically equivalent queries can have different frequencies in a given database instance, although their result tuples are always the same. In particular, although the queries $[\S \wedge \F] \vee [\S \wedge \neg \F]$ and $\S$ are logically equivalent, they have different reference domains: the domain of $[\S \wedge \F] \vee [\S \wedge \neg \F]$ includes also the domain of the query $\F$. This is due to our closed-world assumption: since the entities in the query $\F$ are among those mentioned in the query $[\S \wedge \F] \vee [\S \wedge \neg \F]$, they are included among the {\em potential} answers to the query, although in fact no entity satisfying $\F$ will be an actual answer to the query unless it is also an entity satisfying $\S$.

The final standard property of probability measures on a Boolean algebra is that $P(\X) = 1$, where $X$ is the ``certain event" that contains all possible outcomes. One difficulty with this property from the point of view of our frequency definition is again not so much that the property fails to hold but that it is not straightforward to express. A natural way to translate the axiom into a logical framework is to require that all tautologies or logically necessary queries receive probability 1. For example the query $Student(\A) \vee \neg Student(\A)$ is a tautology when viewed as a logical formula, but it is not a safe query. Another conceptually illuminating difficulty is that in our frequency definition, there is no {\em single} fixed space of possible outcomes or events that is independent of the query being asked. Rather, we define a space of possible outcomes dynamically for every query (i.e., $\dom(\F,\{X_1,...,X_m\})$ for query $\F$). For a {\em given} reference domain, the probability 1 property holds to the extent that we can express it. For example, if the only two possible genders are $\male$ and $\female$, then the query $[Student(\A) \wedge Gender(\A,\male)] \vee [Student(\A) \wedge Gender(\A,\female)$ receives frequency 1 in every database instance.

Finally we show that frequency as defined decreases monotonically with respect to conjunctions. This is important because many algorithms that search for frequent query formulas use this property to avoid exhaustive search. The following
result guarantees that the frequency of a conjunction is less than the
frequency of its conjuncts, which we refer to as the \apriori property.

\begin{proposition}[The \apriori Property] \label{prop:apriori}
Let $\D$
be a database instance with valid ER query $F_{1}$ whose free variables are $\X_1,..,\X_m$ and suppose
that $F_{1}\wedge F_{2}$ is also a valid ER query whose free variables are $\X_1,..,\X_m$. Then
$fr_{\D}(F_{1}\wedge F_{2})\leq fr_{\D}(F_{1})$.
\end{proposition}

\begin{proof}
Clearly $$\tuples(\F_{1} \wedge \F_{2}) \subseteq
\tuples(\F_{1}),$$ and $$\dom(F_{1},\{\X_1,..,\X_m\}) \subseteq \dom(F_{1}
\wedge F_{2}, \{\X_1,..,\X_m\}).$$ So $$\frac{|\tuples(\F_{1} \wedge F_{2})|}{|
\dom(F_{1} \wedge F_{2},\{\X_1,..,\X_m\})|} \leq \frac{|\tuples(\F_{1})|}{|
\dom(F_{1},\{\X_1,..,\X_m\})|}.$$
\end{proof}

{\em Discussion.} Previous approaches to mining multi-relational rules such as \warmr mine rules for just one target table. Our approach in contrast can potentially search the entire space of queries for a given language bias, since by the proposition just established, the a priori property holds for the entire query space, not just for a fixed target table or key atom, given our definition of frequency and support. So compared to an iterative approach where we repeatedly apply a single-table rule miner to different tables in the database, our approach offers computational advantages. Intuitively, our approach combines the results of rule mining for separate tables when it considers rules that involve the separate tables at the same time. For example, suppose that for the $Student$ table, we find that the query $Student(\X) \wedge
Age(\X,30)$ is infrequent. Then from Proposition \ref{prop:apriori} we can conclude that the query $Student(\X) \wedge
Age(\X,30) \wedge Professor(\X)$ is infrequent as well. A traditional single-table rule mining system applied to both target tables 
would have to evaluate this conjunction twice, once with the target table $Student$ and the second time with the target table $Professor$.

The price for the computational advantage of the a priori property holding throughout the query space is that our approach restricts the set of interesting queries compared to an iterative application of single-table rule mining. For example, it may be the case that the rule $Professor(\X) \wedge Student(\X) \rightarrow Age(\X,30)$ receives enough support if evaluated with respect to Professors (because it may be the case that most professors who are also taking courses as students are younger), but does not receive enough support if evaluated with respect to Students (perhaps because very few students are also professors to begin with). Our definition of support based on taking the union of the database tables can be seen as a {\em cautious} approach because if a query is frequent with respect to the union of two tables, it is frequent with respect to either table. So a query that is frequent with respect to the union of the Professor and Student tables is frequent with respect to both.

\section{Conclusion}
The goal of this report was to extend the concept of confidence and support for a new class of association rules which we call entity-relationship rules. Entity-relationship rules are based on the domain relational calculus; they are much more flexible and expressive than standard itemset rules. ER rules allow for negation, nested Boolean combinations, and quantification.The main conceptual contribution of this report is a definition of frequency for entity-relationship queries. Instead of beginning with a specified target table or ``key atom", we dynamically define a reference or base domain of individuals for each ER query. The key idea of our definition is to take the base set of entities of a conjunctive query to be the union of the conjuncts' base sets. For example, the frequency of the query $Professor(\X) \wedge Customer(\X)$ is computed with respect to the union of Professors and Customers.
We proved that our frequency definition satisfies standard axioms for probabilities and validates the \apriori property: the frequency of a conjunction is no greater than the frequency of any conjunct.

As usual in data mining, there is a tradeoff between the expressiveness of the rule or pattern language, and the difficulty of searching for significant patterns. Our rule language is very general and in practice a computational search for interesting entity-relationship rules will require a language restriction (bias). A central topic for future research is to explore language restrictions that make feasible a computational search for interesting entity-relationship rules.

\section*{Acknowledgements}
This research was supported by Discovery Grants to the first and third author from the Natural Sciences and Engineering Council of Canada.

\end{document}